\documentclass[12pt]{article}
\usepackage{graphicx}
\begin{document}

\centerline{\large Experimental realities refuting existence of p=0 condensate} 

\centerline{\large in a system of interacting bosons : I. Electron bubble}

\vspace{2cm}

\centerline{\bf Yatendra S. Jain}

\bigskip
\centerline{\bf Department of Physics}

\smallskip
\centerline{\bf North-Eastern Hill University, Shillong - 793022, India}

\vspace{2.5cm}
\begin{abstract}

Physical reality of the existence of electron bubble in liquid 
$^4He$ (or $^3He$) renders a {\it clear experimental evidence} 
for a quantum particle (in an interacting environment as seen 
by electron in liquid helium) to occupy exclusively a space of 
size $\lambda/2$ that, obviously, depends on its energy/momentum.  
This unequivocally proves that {\it no particle} in a system of 
interacting bosons such as liquid $^4He$ has momentum $p=0$; in 
stead, {\it all particles} in the ground state of such a system 
are in the single quantum state of energy $\varepsilon_o = h^2/8md^2$ 
or momentum $p = h/2d$.

\end{abstract}



\vspace{2.5cm}

\noindent
\copyright \, \, by authors
\newpage

\bigskip
Superfluidity and related properties of the low temperature 
phase of a {\it system of interacting bosons} (SIB) such as 
liquid $^4He$ [1] and trapped dilute gases [2] are, conventionally, 
attributed to an assumed existence of $p=0$ condensate, -a 
macroscopic fraction [$n_{p=0}= N_{p=0}/N$] of its total number 
of particles, $N$, occupying zero momentum ($p=0$) state of a 
single particle confined to its volume, $V$; remaining 
particles, $N - N_{p=0}$, believed to occupy different 
states of non-zero momenta, $k_1$, $k_2$, $k_3$,  .... {\it etc.}, 
(in wave number unit) form non-condensate component.  
This follows from the statistical analysis of non-interacting 
bosons [3] and Bogoliubov's field theoretical study of a system 
of weakly interacting bosons [4].  Analyzing this distribution 
(believed to be the origin of superfluidity and related aspects 
of a SIB for about seven decades now), we recently discovered [5] 
that it does not correspond to lowest possible energy and, as such, 
it is unacceptable for the ground state (G-state) of any physical 
system. Our analysis [5] also reveals that in the 
state of least possible energy, needed for the G-state, of a SIB 
all particles have identically equal energy 
$\varepsilon_o = h^2/8md^2$ (equivalent momentum $q_o  = \pi/d$) 
which represents the G-state energy (momentum) of a particle 
trapped in a cavity formed by its nearest neighbors and the 
state is shown to have least possible energy.  Consequently 
not even a single particle is expected to have $p=0$.   In 
other words, our analysis reveals that the laws of nature, which 
define the G-state of a physical system to have minimum possible 
energy, forbid the existence of $p=0$ condensate and the said 
non-condensate in the G-state of a SIB.  In this paper, we use 
the physical reality of the existence of an {\it electron bubble} 
(EB) [6,7] in liquid helium to place these inferences on 
unshakable foundation. 

\bigskip
An excess electron in liquid helium exclusively occupies a self 
created spherical cavity (known as electron bubble) of certain 
radius when it assumes its {\it lowest possible energy} in the 
cavity; to create the said cavity, electron exerts its zero-point 
force on helium atoms in its surroundings and works against the 
forces originating from inter-atomic interactions and external 
pressure on the liquid [6,7].  It is evident that the bubble 
formation is a consequence of the facts that: (i) the electron 
experiences a strong short range repulsion with helium atoms which 
forbids its binding with them and (ii) an electron is a quantum 
particle which behaves as a wave packet whose size, 
$\lambda/2 = h/p = h/\sqrt{(2mE)}$, increases with the decrease 
in its energy E (or equivalent momentum $p$).  This implies that any 
quantum particle that experiences similar repulsion with helium 
atoms should have similar state in liquid helium and this is found 
to be true with positron [8] and other particles (ions, atoms, 
molecules, etc. [9]). 

\bigskip
Since $^4He$ atom is also a quantum particle and it also has a 
short range strong repulsion with other $^4He$ atoms, each $^4He$ 
atom in its lowest possible energy in liquid $^4He$ should 
exclusively occupy a spherical cavity with an energy $h^2/8mv^{2/3}$ 
where $v$ is the volume of the cavity.  Assuming that $i-$ atom occupies 
$v_i$ volume we can express the total sum of the G-state energy 
of $N$ particles in the liquid as 
$$E = \sum_i^N\frac{h^2}{8mv_i^{2/3}} \eqno(1))$$ 

\noindent
with
$${\rm V}  = \sum_i^N v_i \eqno(2)$$

\noindent
A simple algebra reveals that $E$ has its minimum value only and 
only when all $v_i$ are identically equal to ${\rm V}/N = d^3$.  This 
concludes that the right value of the G-state energy of a SIB like 
liquid $^4He$ and trapped dilute gases should be 
$$E_o = N\frac{h^2}{8md^2} = N\varepsilon_o \eqno(3)$$ 

\noindent
which clearly proves that no particle in the G-state of a SIB 
has zero energy/momentum. Evidently, the laws of nature, 
responsible for the physical existence of EB, forbid the 
existence of even a single particle to have $p < h/2d$; the 
question of a particle having $p=0$ or a macroscopic fraction 
of particles in a SIB having $p=0$ does not arise. 

\bigskip
The reasons, for which an electron in liquid $^4He$ or $^3He$ 
exclusively occupies a large size spherical cavity 
({\it electron bubble}) are, obviously, applicable to each $^4He$ 
atom in liquid $^4He$ which should occupy exclusively a 
spherical cavity with certain energy $\varepsilon_o = h^2/8md^2$ 
or equivalent momentum $h/2d$.  Thus the physical reality of the 
existence of an EB in liquid $^4He$ and $^3He$ unequivocally 
establishes that no $^4He$ atom in the G-state of liquid $^4He$ 
has $p=0$ confirming that the nature forbids $p=0$ condensate of 
helium atoms in the ground state of liquid $^4He$.  This 
conclusion, obviously holds true for the atoms in Bose Einstein 
condensate of trapped dilute gases.  In fact no particle of such 
gases has zero momentum because the wave mechanics of a particle 
in harmonic trap has to have a least energy of $(3/2)h\nu$ 
(where $\nu$ is the fundamental frequency of oscillations of the particle 
in the trap).  While we find that Eqns.(1-3) are consistent with 
excluded volume condition [10] which states that each particle 
such as $^4He$ atom occupies certain volume {\it exclusively} due 
to its short range hard core interaction with neighboring 
particles, it is important to note that Kleban [10], in his paper 
published long back in 1973, concluded that $p=0$ condensate 
contradicts the excluded volume condition.  However, his inference 
was not given due importance for so long for an undue bias in favor 
$p=0$ condensate as the origin of superfluidity and related 
properties of a SIB.
    
\bigskip
In what follows, we compare the momentum distribution of particles 
in the G-state of a system of non-interacting bosons [Fig 1(A)], and 
the momentum distributions of bosons in the G-state of a SIB as found by 
Bogoliubov [4] with that concluded by our studies [5], respectively, 
depicted in Fig.1(B) and Fig.1(C), for their better perception.  It 
is clear that $100\%$ particles in a SIB at absolute zero occupy a 
state of non-zero momentum, $p=h/2d$ as depicted in Fig.1(C).  
Although, the existence of an electron bubble clearly proving that a 
quantum particle in its G-state exclusively occupies certain space 
of the size, $\lambda/2$, of its wave packet particularly when it 
interacts with its neighbors through a strong repulsive force, it is 
unfortunate that this reality was never used to determine the G-state 
of liquid $^4He$, liquid $^3He$ and electron fluid in conductors.  
Had it been so, our understanding of superfluidity and superconductivity 
would have been as simple as what is concluded in [11, 12] and the G-state of 
a SIB [as depicted by Fig.1(C)] could have been concluded long back.

\newpage
 \bigskip
\begin{figure}
\begin{center}
\includegraphics[angle = 0, width =.6\textwidth]{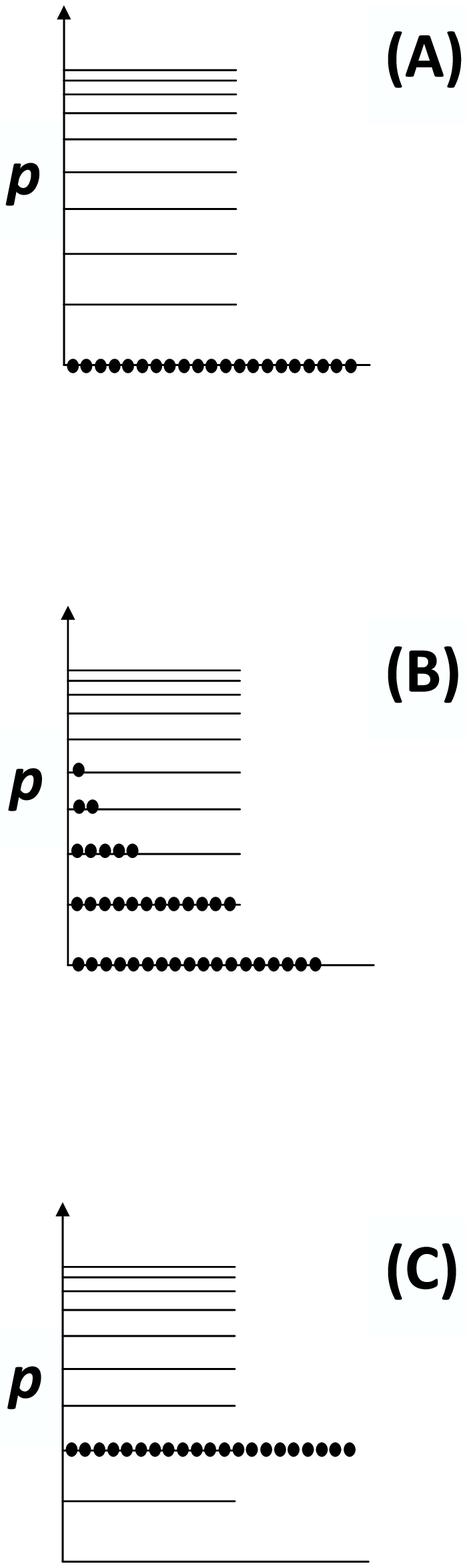}

\end{center}
\bigskip
\noindent
Fig.1 : Schematic of distribution of $N$ bosons in their 
ground state. (A) All the $N$ particles occupy $p=0$ state 
in a system of non-interacting bosons, (B) depletion of 
$p=0$ condensate ({\it i.e.} only a fraction $n_{p=0} = N_{p=0}/N$ 
of $N$ bosons occupy $p=0$ state) in weakly interacting boson 
system as predicted 
by Bogoliubov model [4], and (C) all the $N$ particles occupy 
a state of $p= p_o = \hbar q_o = h/2d$ and $\hbar K=0$ as 
concluded by this study and our recent analysis [5]. 

\end{figure}

\bigskip
\begin{thebibliography}{??}
\bibitem[1]{Kn:gnus}
C. Enss, S. Hunklinger, Low Temperature Physics, Springer Verlag, Berlin, 
2005.
\bibitem[2]{Kn:gnus}
F. Dalfovo, S. Giorgini, L.P. Pitaevskii, and S. Stringari,  
Rev. Mod. Phys. 71, 463-512 (1999), and Arxiv/Cond-mat/9806038.
\bibitem[3]{Kn:gnus}
R.K. Pathria,  Statistical Mechanics,
Pergamon Press Oxford (1976). 

\bibitem[4]{Kn:gnus}
N.N. Bogoliubov, J. Phys, USSR, 11, 23-32 (1947).

\bibitem[5]{Kn:gnus}
Y.S. Jain, {\it The p=0 condensate is a myth}, arxiv:cond-mat/1008.240v2 
\bibitem[6]{Kn:gnus}
M. Rosenblit, J. Jortner, Phys. Rev. Lett. 75, 4079-4082 (1995);
M. Farnik, U. Henne, B. Samelin, and J.P. Toennies,  
Phys. Rev. Lett. 81, 3892-3895 (1998)
\bibitem[7]{Kn:gnus}
H. Marris, and S. Balibar, Phys. To-day 53, 29-34 (2000)
\bibitem[8]{Kn:gnus}
P. Hautojarvi, M.T. Loponen, and K. Rytsola, J. Phys. Atom. Mole. Phys. 
9, 411-422 (1976)

\bibitem[9]{Kn:gnus}
J.P. Toennies, and A.F. Vilesov, Ann. Rev. Phys. Chem. 49, 1-41 (1998)


\bibitem[10]{Kn:gnus}
P. Kleban,  Phys. Lett. A 49, 19-20 (1973).

\bibitem[11]{Kn:gnus}
(a) Y. S. Jain, Ind. J. Phys. 79, 1009-14 (2005); the factor 
$|\sin{({\bf k}.{\bf r}/2)}|$ in Eqn.5 for $\Psi^+$ in this paper 
should be read as  $\sin{|({\bf k}.{\bf r}/2)|}$ and $E_g(T)/Nk_BT$ 
in Eqn.25 should be read as $E_g(T)/Nk_B$
    
(b) Y.S. Jain, Macro-orbitals and microscopic theory of a system 
of interacting bosons, arXiv:cond-mat/0606571

\bibitem[12]{Kn:gnus}
Y.S. Jain, Basic foundations of the microscopic theory
of super conductivity, arXiv:cond-mat/0603784, (2006).



\end{thebibliography}
\end{document}